# Flow Characteristics and Cores of Complex Network and Multiplex Type Systems

Olexandr Polishchuk

Laboratory of Modelling and Optimization of Complex Systems

Pidstryhach Institute for Applied Problems of Mechanics and Mathematics National Academy of Sciences of Ukraine

Lviv, Ukraine

od_polishchuk@ukr.net

**Abstract—Subject of research is complex networks and network systems. The network system is defined as a complex network in which the flows are moved. Classification of flows in the network is carried out on the basis of ordering and continuity. It is shown that complex networks with different types of flows generate various network systems. Flow analogues of the basic concepts of the theory of complex networks are introduced and the main problems of theory of complex networks in terms of flow characteristics are formulated. Local and global flow characteristics of networks bring closer the theory of complex networks to the systems theory and systems analysis. Concept of flow core of network system is introduced and defined how it simplifies the process of its investigation. Concepts of kernel and flow core of multiplex are determined. Features of operation of multiplex-type systems are analyzed.**

*Keywords—complex network; network system; flow; multiplex; core; kernel*

## I. Introduction

The study of any system begins with defining its composition and structure [1, 2]. Cognition of the laws of its functioning and interaction with the environment should follow further [3]. Systems of various types and purposes are the most difficult subject of such investigation [4, 5]. This is explained by the need to determine the structure, process of operation and interaction of large number of different objects that act to achieve the common and often not fully understandable for the researcher goal. A number of disciplines engaged in the study of complex systems [6-10] (general systems theory, systems analysis, operations research, systems engineering, mathematical modeling and optimization, etc.). In each of the natural and human sciences can be distinguished certain "system" subsection, which deals with its subject area [11-14]. In order to understand the structure and the laws of functioning of the large complex system is need a holistic view of it.

An attempt to comprehend a complex system as a whole at least at the level of analysis of its structure was the main reason for creation of theory of complex networks (TCN) [15, 16]. Network structures are in the macro- [17] and micro-world [13]. They are the most widespread structures in biological systems (neuronal, protein, metabolic, food, ecological networks etc.) [18-21]. A such structures are common in the human society (economic, social, financial, political, religious, professional, family, and many others) [22-24].

Sometimes complex networks (CN) call the systems [25]. However, let us imagine a railway network as a set of stations and railway tracks connecting them where the railway traffic is absent or a gas or oil supply network as a set of compressor stations and pipelines connecting them which are not pumped with gas or oil. How would computer network represented as a set of servers, computers and wire or wireless connection means look like, if it wasn't used for exchange of information? That is, the complex networks actually only reflect the structure, in other words, are the "frame" of the majority of real complex systems. The last statement does not diminish the value of TCN and the problems that are solved by it. After all, the reason of inefficient operation of many real-world systems is the shortcomings of their structures. With the other hand, have existed and exist the systems that can be investigated only by studying their structures. Among them are network structures of the universe [26], the study of which allows us to find the solution of many cosmological problems. Archaeologists are finding the infrastructure networks of lost civilizations, paleontologists - the habitats of extinct animals and plants. At one time these were quite large and complex systems, which we are now able to investigate solely on the basis of their fossil "network" remains.

The main goal of creation and development of the most known networks is to provide the movement of a certain type of flows. Of course, the motion of flows is not fully reflects all processes in complex systems. However, these processes are the way to implement functions that contribute to achieving goal. Namely, network nodes carry out the processing of flows (creation, reception, maintenance) [27], and the edges ensure their smooth passage. That is, a complex network system is not just a set of components connected in a certain structure, but the process of co-operation of these components in order to achieve the goal - to ensure the movement of flows. Development of the theory of network flows (TNF) began in the mid-fifties of the last century [28]. This theory solves a number of practically important problems relating to the network as a structure to ensure the movement of flows. Among such problems include [29] the maximum flow problem and related minimum cut problem, the matching problem and its generalization the assignment problem, the maximum dynamic and steepest flow problems, generalization of which is the evacuation problem. Solving these problems is certainly important to optimize the operation of network systems (NS). However, they mostly have to deal with the structural features of the system than to its operation process. Thus, TCN and TNF mainly investigate the properties of structure of network system, whereas the main task of the system disciplines above all is a study of functions that are implemented by components of this structure, and allow us to achieve the set goal.

The purpose of this article is to identify ways to approximation of structural and functional directions of the study of complex network systems. Two ways of such approximation can be distinguished. The first way is to introduce the basic concepts of systems disciplines in the conceptual apparatus of TCN. As will be shown below, this way comes up against certain difficulties that are associated with ambiguity of mapping "complex network → network system". The second way, under choice of which this mapping is unique, is to determine the flow characteristics of network systems as an alternative to structural characteristics of complex networks. Flow characteristics allow us to show that the equivalent from viewpoint of TCN network elements can play a different role in the operation of network system. Herewith the content of formulated in terms of flow characteristics of the main problems of TCN essentially extends and deepens.

## II. COMPLEX NETWORKS AND NETWORK SYSTEMS

In general, an arbitrary network is defined as a statistical assembly, i.e. a set of networks with each network having certain probability of implementation, or as a set of all possible states of the given network [25]. On the other hand, complex networks are graphs, i.e. the sets of nodes connected by some edges with nontrivial topological properties. When talking about real networks these properties determine network operation features [15]. Among the basic concepts of TCN should be called orientation of network, local and global structural characteristics of nodes and links, common topological properties of network substructures (cliques, communities, etc.), and the network as a whole. The subject of research in TCN is [16] the creation of universal network models, definition of statistical properties that characterize their behavior as well as the forecasting of behavior of networks when changing their structural properties.

In every city there are systems of electrical, gas and water supply, fixed telephone, cable television, internet and postal services, etc. All these systems have practically identical network structure in which most of nodes constitute individual houses and apartments. However, despite the identity of structures, these are significantly different systems. It is argued [22, 24] that the natural and artificial, physical and biological networks, networks of micro- and macrocosm have many similarities. The human and chimpanzee genomes coincide at 95% [30]. However, this does not make the chimpanzee homo sapiens. Similarity of structures helps in the development of universal methods of investigation of these structures, but not always the systems.

System disciplines operate with such concepts as the goal, function, state, process, behavior, stability, controllability, etc [31]. The subject of their research is the study of different classes and types of systems, basic laws of its behavior, process of goal formation, functioning, development and interaction with the environment [7, 32]. The first step in the study of any system is identification and investigation its composition and structure. In the case of network systems this step can be carried to the prerogative of TCN. Further follows the definition of goals of the system existence and operation, implementation of which contributes to the achievement of these goals. In general in the systems disciplines the structure of system is considered together with the functions that are implemented by the components of this structure and the system at a whole, at that the function has priority over the structure [10]. Thus, both conceptual apparatus of TCN and systems disciplines and subjects of their research practically almost no overlap.

The flows in network can be continuous (e.g. power resources), discrete (e.g. trains) or continuous-discrete (motor transport flows). The motion of flows in the network can be ordered, partially ordered or unordered. An ordered we call motion of flows that move in accordance with a certain schedule. Movement of trains in the railway system is such. Partially ordered we call the motion of flows, part of which moves according to a certain schedule. The movement of communal transport in cities, the suburban and interurban bus service is examples of such movement. We call unordered the motion of flows, which is difficult to describe using the deterministic regularities. Information flows in the Internet are an example of such movement. Networks with different types and levels of flow ordering generate the different network systems [33]. Classification of flows in the network can be deepened on the basis of their periodicity, uniformity, controllability, material carrier, etc. This allows us to accomplish a more accurate classification of network systems.

Consider the series of basic concepts of TCN and show how the account of motion of flows in the network can significantly enrich their content and solve some important problems.

### III. FLOW CHARACTERISTICS OF NETWORK SYSTEMS

Every network node may be connected with other nodes by some number of edges. The network is called directed if these connections have a direction. If all edges between nodes are symmetrical, the network is called an undirected. Orientation of network is well described by the adjacency matrix. We assume that between any two nodes of network can be only one connection and there are no so-called "loops". Then, if the network has $N$ nodes, the elements $a_{ij}$ of adjacency matrix $A = \{a_{ij}\}_{i,j=1}^{N}$ are equal to 1, if the connection between nodes $n_i$ and $n_j$ exists and are equal to 0 if there is no such connection. Obviously, the adjacency matrix is symmetric for undirected networks. Many properties of CN may be expressed using the adjacency matrix. For example, such important concept of TCN, as the degree $d_i$ of node $n_i$, for undirected networks is determined by the elements of adjacency matrix as follows: $d_i = \sum_{j=1}^{N} a_{ij}$. For directed networks are used such concepts as an output degree $d_i^{out}$ and an input degree $d_i^{in}$ of the node $n_i$. The output degree of node determines the number of connections that are "go" of it, and the input degree – the number of connections that are "come" in the node. It is easy to see that

$$d_i^{out} = \sum_{j=1}^{N} a_{ij}, \quad d_i^{in} = \sum_{j=1}^{N} a_{ji},$$

and for undirected networks $d_i^{out} = d_i^{in}$.

Assume that flows are continuously distributed along the edges of CN and nodes $n_i$ and $n_j$ are connected by edge (curve) $y_{ij}(x)$, $x \in [x_i, x_j]$. Denote $\rho_{ij}(t, x, y_{ij}(x))$ – the density of flow which moves from node $n_i$ to node $n_j$ at the point $(x, y_{ij}(x))$, $x \in [x_i, x_j]$, in moment of time $t \in T_l = [t_l, t_{l+1}]$, $l = 0,1,2,...$, $t_0 = 0$. Then the total volume of flow directed from node $n_i$ to node $n_j$ at time $t$ on the edge $(n_i, n_j)$ is determined by the ratio

$$v_{ij}(t) = \int_{(n_i, n_j)} \rho_{ij}(t, x, y_{ij}(x)) dx = \int_{x_i}^{x_j} \rho_{ij}(t, x, y_{ij}(x)) \sqrt{1 + (y_{ij}'(x))^2} dx,$$

and volume of flow which passes from node $n_i$ to node $n_j$ during period $T_l$, $l = 0,1,2,...$, is determined by the ratio

$$V_{ij}^{(l)} = \int_{t_l}^{t_{l+1}} v_{ij}(t) dt.$$

Usually the presence and volume of flows that move by the network edges is defined simply enough.

Define flow adjacency matrix $B_f^{(l)} = \{b_{ij}^{(l)}\}_{i,j=1}^{N}$ of the network system during period $T_l$ by the ratio

$$b_{ij}^{(l)} = V_{ij}^{(l)} / \max_{m,k=\overline{1,N}} V_{mk}^{(l)}, \ i,j = \overline{1,N}.$$

The elements of matrix $B_f^{(l)}$ determine flow connection strength between network nodes during period $T_l$, $l = 0,1,2,...$ With the change of values $l$ the matrix $B_f^{(l)}$ represents the dynamics of changes in movement of flows in network over time. Obviously, the flow adjacency matrix both for directed and undirected networks is non-symmetric.

Define the flow degree of node $n_i$ of network system. Denote the output flow degree of node $n_i$ as $f_{i,l}^{out} = \sum_{j=1}^{N} b_{ij}^{(l)}$ and the input flow degree of the node $n_i$ as $f_{i,l}^{in} = \sum_{j=1}^{N} b_{ji}^{(l)}$. Then the total flow degree of node $n_i$ is determined by the ratio $f_{i,l} = f_{i,l}^{out} + f_{i,l}^{in}$, $f_{i,l} = f_{i,l}^{out} + f_{i,l}^{in}$. It is obvious that the content of adopted in TCN concept of degree of node and the above-defined concept of flow degree of node are significantly different. However, we use the same term "degree" because of the similarity of method of their determination by means of elements of corresponding adjacency matrices. Further, in order to avoid ambiguity, we shall call structural the determined in TCN term of degree of node.

From the viewpoint of TCN all nodes with the same structural degree are equally important for the system. Flow degrees contribute to functional ordering of nodes priority with equal structural degrees. Moreover, they confirm that the nodes with the lesser structural degree, but larger flow degree are more important in the system operation. Therefore, hereinafter all flow concepts we shall assume the characteristics of not so a complex network as a network system.

In TCN is usually accepted that the connection strength between the nodes is equal to 1 if there is an edge between them and to 0 if there is no such edge. Thus, the connection strength between any nodes of connected network is also formally equal to 1 (network is called a connected if there is a path between its two arbitrary nodes). In the case of disconnected networks the connection strength between two arbitrary nodes is equal to 1 if there is a path between them, and is 0 if there is no such path (nodes belong to disconnected network components). Obviously, the flow characteristics allow us to create a more adequate picture of connection strength between network nodes. Namely, we will determine the flow connection strength between two arbitrary nodes of connected network, as the maximal from all minimal values of flow adjacency matrix, which lie on the paths connecting these nodes. The connection strength is used during the solution of many problems of modeling of complex networks (network synchronization, its controllability, observability, etc.) [34-36]. Delay of flow at node $n_i$, directed through a series of intermediary nodes to the node $n_j$, delays the receipt of this flow in receiving node. Thus, the flow connection strength between remote network nodes determines the possibility of mediated influence of one node to another. In this case remote node can affect more than the node with which there is a direct connection (edge). The connection strength between network nodes, identified by means of its flow characteristics, determine the quantitative expression of interaction between the elements of network system.

The degree of node is a local characteristic of CN. Among global network characteristics can be called betweenness centrality of node $n_i$. In TCN this characteristic defines [15] as the total number of shortest paths between all other nodes that pass through the node $n_i$ and reflects the topology of network as a whole. This parameter reflects the importance of node in network structure. We propose to define the flow importance of node in the system taking into account the volume of flows, which pass through this node in comparison with other network nodes.

The flow importance $\omega_i$ of node $n_i$ during period $T_l$ is determined by the ratio

$$\omega_{i,l} = f_{i,l} / \max_{j=1,N} f_{j,l}.$$

In TCN the main attention focuses on the definition of properties of network nodes. Accounting of movement of flows allows us to determine not only the importance of network nodes but also the load of its edges. The load of edge $\theta_{ij}$ that connects the nodes $n_i$ and $n_j$ during period $T_l$ is determined by the ratio

$$\theta_{ij}^{(l)} = b_{ij}^{(l)} / \max_{m,k=1,N} b_{mk}^{(l)}.$$

Increase or decrease the flow importance of nodes and load of edges, which belong to some connected subnetwork of source network may be indicative of activation (acceleration of development) of corresponding system component, or vice versa – the distribution of depressive processes in this component.

IV. PROBLEMS OF SEARCH OF FICTIVE AND HIDDEN NODES AND EDGES

Often the Internet user is registering on the certain site, i.e. establishes a connection, but never visits this site later. Such connection can be considered as fictive, i.e. does not really existing. If all connections of the node with other nodes of CN are fictive, this node also is considered as fictive. The flow adjacency matrix allows us to determine in structure of source network (Fig. 1a) the fictive nodes and edges (Fig. 1b) and remove them (Fig. 1c).

Between networks nodes can move flows, although the connection between them in network structure is not formally established. Accounting of motion of flows allows us to detect such hidden connections. The flow can be sent from one node to another, but in the process of movement it can be "split" and come in several nodes. This means that there is a hidden node, which redistributes the motion of flows. Note that using TCN the problem of detection of fictive and hidden nodes and edges is practically impossible to solve. Further we will assume that the structure of network includes only the real nodes and edges (Fig. 1d).

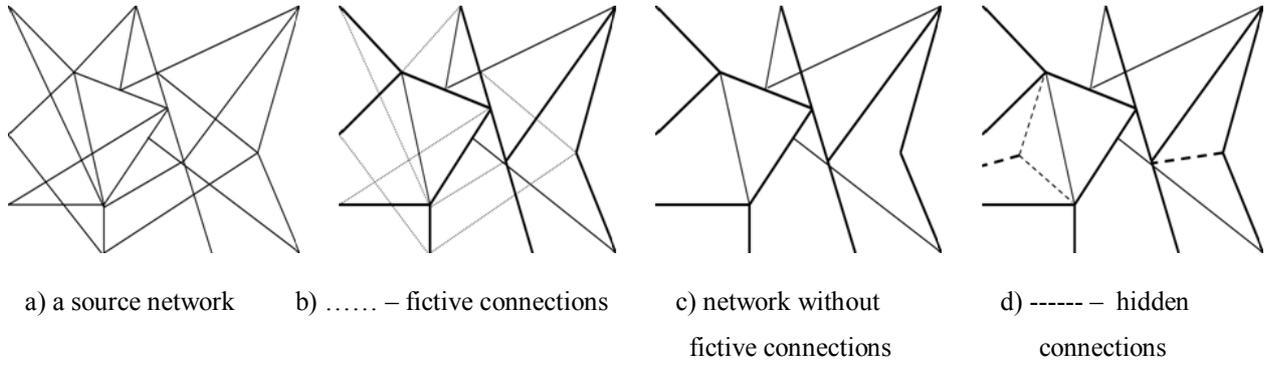

a) a source network   b) ...... – fictive connections   c) network without fictive connections   d) ------ – hidden connections

Fig.1. Fragment of complex network with fictive and hidden connections

Volume of flow, which emerges from node $n_i$ in direction of node $n_j$ during period $T_l$ is determined by the ratio

$$w^{(l)}_{ij,out} = \int_{t_l}^{t_{l+1}} \rho_{ij}(t, x(n_i), y_{ij}(x(n_i)))dt.$$

Assume that flow moves along the edge $(n_i, n_j)$ from node $n_i$ towards node $n_j$ at an average rate $u_{ij}$ and $l_{ij}$ is the length of edge $(n_i, n_j)$. Then the volume of flow, which should arrive in the node $n_j$ from node $n_i$ during period $[t_l + l_{ij}/u_{ij}, t_{l+1} + l_{ij}/u_{ij}]$ is determined by the ratio

$$w^{(l)}_{ij,in} = \int_{t_l}^{t_{l+1}} \rho_{ij}(t + \frac{l_{ij}}{u_{ij}}, x(n_j), y_{ij}(x(n_j)))dt.$$

Losses or increase of the volume of flows that come out from node $n_i$ and must arrive at the node $n_j$, can be analyzed by comparing the values $w^{(l)}_{ij,out}$ and $w^{(l)}_{ij,in}$. The discrepancy between these values may indicate the presence of hidden nodes on the edge $(n_i, n_j)$, which are removed or added some volume of flows. In general $\left| w^{(l)}_{ij,out} - w^{(l)}_{ij,in} \right| \leq \delta^{(l)}_{ij}$, where $\delta^{(l)}_{ij}$ defines the natural loss or gain (e.g. an increase or decrease in humidity of the transported product) during the movement of flows.

Assume that $W^{(l)}_{i,in}$ is the total volume of flows that arrive during the period $T_l$ in the node $n_i$. Denote by $J_i = \{j_1, j_2, ..., j_i\}$ the set of numbers of nodes connected with $n_i$. Then fulfillment of inequality

$$W^{(l)}_{i,in} > \sum_{j \in J_i} w^{(l)}_{ji,in}$$

indicates the presence of unregistered connections of node $n_i$ through which flows are coming in. If the country has a large number of registered nonworking enterprises and a big shadow sector, enterprises of which works, but do not pay taxes, it significantly distorts the picture of real economical state. Summing up it can be said that accounting of movement of flows allows by means of expulsion from network composition the fictive nodes and edges, and the inclusion of hidden elements to make a more precise view about the structure of CN and determine the prospects of its development.

V. PROBLEMS OF MODELLING OF COMPLEX NETWORKS AND NETWORK SYSTEMS

Above we showed how accounting of flows motion allows us to determine the real structure of network. However, there is still a lot of modelling problems for CN and NS. One of the first is called the problem of finding alternative paths of flows motion. This problem occurs on a daily basis during the peak hours in the transport networks of large cities. Traffic jams on the main highways are forcing many drivers look for alternative paths of movement. In general, this is a particular example of more global problem. It consists in a temporary withdrawal from network structure of some its subnetwork. In Fig. 2a is schematically reflected a fragment of source network, part of which became unavailable (a gray area in Fig. 2b). As a result, we get frequently a disconnected network (Fig. 2c). Such situations are caused by epidemics (introduction of quarantine during Ebola virus in Africa), natural disasters, regional military conflicts, embargo in respect of certain countries, etc. Here, together with the task of finding alternative paths of flows motion arise problems of replacement nodes-generators (for example, manufacturers of some products) and nodes-recipients (product consumers). As a result of earthquake in Japan on March 11, 2011 in the area of Fokusimy were destroyed nearly 40 enterprises - manufacturers of components for the automotive industry. Because of this, almost all Japanese automakers temporarily stopped their conveyors. That is, the unavailability of some subnetwork of source network often resulting in destabilization of the whole system operation.

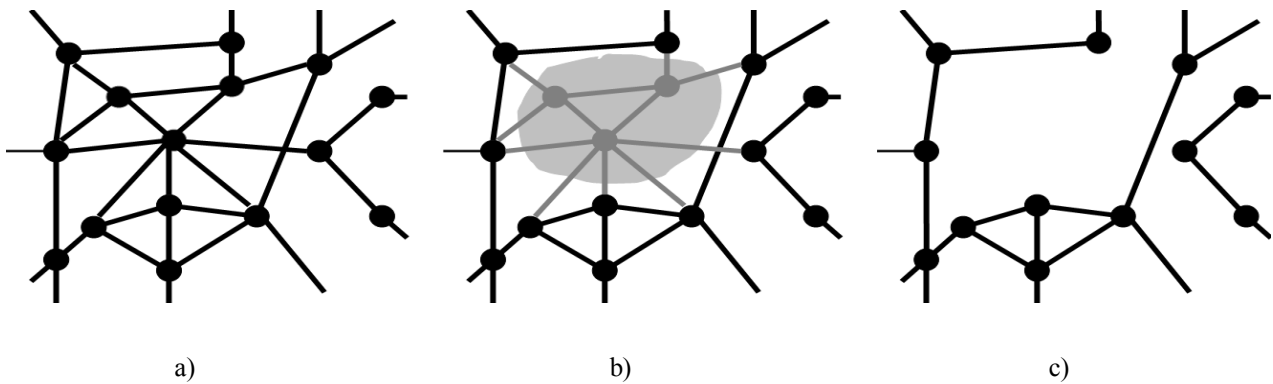

a)            b)            c)

Fig. 2. Problems of search of alternative paths

In TCN solution of the problem of search of alternative paths of flows motion consist in finding at least one of the shortest ways to circumvent unavailable network component. In studying of NS we encounter to the problem of redistribution of flows movement. Usually redirect all flows on one path is not possible. This is due to its limited handling capacity. In the majority network systems separate flows have different priorities. This priority may be determined by the type of flow (international trains have higher priority than suburban), its purpose, importance of generator or recipient of flow etc. The basic principle of flows redistribution is as follows [27]: the higher is priority of flow, the shorter path or delay of its movement. That is, the highest priority flows are directed along the shortest path is not yet depleted its handling capacity. With decreasing of priority flows are redirected by other (longer) paths or delayed.

Among other important issues of modeling of complex networks and network systems can be called problems of synchronization, controllability and observability [34-36]. The formulation and solution of these problems for CN and NS differ substantially in content and complexity. Any mode of semaphore work that does not create accident situations during the movement of vehicles synchronizes the network one way or another. However, this mode does not necessarily provide the synchronization of traffic flows on the network at different times of day, when the direction and intensity of the traffic is changing. In other words, the motion of flows in network is dynamic process. Herewith synchronization of network is a necessary but not sufficient condition for synchronization of network system. Motion of flows in the network will be considered synchronized if the total time of its delay during processing in the nodes and passing through the edges is equal to zero. This means that the synchronized movement with the time ceases to be such and requires new synchronization. It has been established [34], that one of the main parameters that influence on network synchronization is the connection strength between its nodes. The accuracy of determination of this strength is not less important. Use in known models of network synchronization [37-39] the defined above flow connection strength and level of ordering of flows movement in system can significantly improve these models. In the case of continuous flows in CN, proposed model of network synchronization will have the next form

$$\frac{\partial \theta_i}{\partial t} = \omega_i + (1-\alpha)\zeta_i(t) + \frac{1}{N}\sum_{j=1}^{N} b_{ij} \sin(\theta_j - \theta_i),$$

where $\theta_i$ is phase of $i$-th network node, $\omega_i$ is eigenfrequency of $i$-th network node, $\alpha$ is the level of ordering of flows movement in the network, $\zeta_i$ is the level of noise in the $i$-th node, $b_{ij}$ are the elements of flow adjacency matrix, $t \in [0,T]$. The value $\alpha$ is determined by the ratio $\alpha = \frac{V_{or}}{V_{net}}$, where $V_{or}$ is volume of ordered flows, $V_{net}$ is total volume of flows in NS. In proposed by J. Gomez-Gardennes [39] network synchronization model is adopted that $\alpha=1$ and $b_{ij} = a_{ij} = const$ (in Y. Kuramoto model [37] $a_{ij}=1$), $i,j = \overline{1,N}$. H. Daido [38] researched models with the noise without accounting the level of ordering of flows movement. In proposed above model, the connection strength between the nodes has the objective content and noises are suppressed by ordering of motion.

Under controllability in general understands the possibility of transfer the system from given initial state to given final state for a finite period of time. Now the problems of controllability of complex networks and network systems are in the process of comprehension and formulation. Currently these tasks are limited by the simplest models of oriented dense and homogeneous networks. It has been established [35], that CN can be controlled using a few high-degree driver nodes. Among the most important problems tied with the issue of controllability of network systems can be called a fight or, on the contrary, facilitation the spread of so called cascade effects [40, 41], redirect of flows at the alternative paths, dynamic synchronization of flows motion, quick adaptation to changes in motion schedule in the systems with fully or partially ordered movement, expansion or narrowing of the network depending on the stage of its life cycle, etc.

VI. CORES OF COMPLEX NETWORKS AND NETWORK SYSTEMS

During the study of large network systems the problem of dimension of their models arises. Complex networks can have millions and billions elements (nodes and edges connecting them). The number of processes that occur in the network systems can be much greater [42]. The main way to solve this problem is to simplify the model by discarding elements, the least important from researcher's point of view [43]. However, during such simplification the problem of preservation of model adequacy arises. We have considered above one of the options for reducing of dimension of the model without losing its adequacy which consists in the removal from structure of NS of fictive nodes and edges. Another way to simplify the models of large complex networks is the introduction of concept of network $k$-core, i.e. the largest subnetwork of source network all nodes of which have the structural degree not less than $k$ [44, 45]. The use of flow characteristics of NS allows us to introduce the concept of flow core of network system. Flow cores of NS make it possible to develop the more adequate from a functional point of view simplified network models than their $k$-cores.

Assume that the network includes $N$ nodes. Let us denote $B_f^* = \{b_{ij}^*\}_{i,j=1}^N$, where $b_{ij}^* = b_{ji}^* = (b_{ij} + b_{ji})/2$, $b_{ij}^* \in [0,1]$, $i,j = \overline{1,N}$. We determine the flow $\lambda$-core of NS as the greatest subnetwork of its structure in which all elements of matrix $B_f^*$ have the values not less than $\lambda$, $\lambda \in [0,1]$. In Fig. 3a is schematically reflected a small fragment of real transport network of large city. In Fig. 3b is contained the same fragment with a mapping of volume of flow movement, the value of which is proportional to the thickness of lines. In Fig. 3c and 3d are showed 4-core of CN fragment and flow 0.9-core of NS fragment, respectively. Obviously, the flow $\lambda$-core of NS gives more important information for system research than $k$-core of its structure. So, disconnect of node B with the structural degree 3, which lies on the path of intensive flow movement will lead to significantly greater problems in their redistribution by other ways than the disconnect of node A with the structural degree 4.

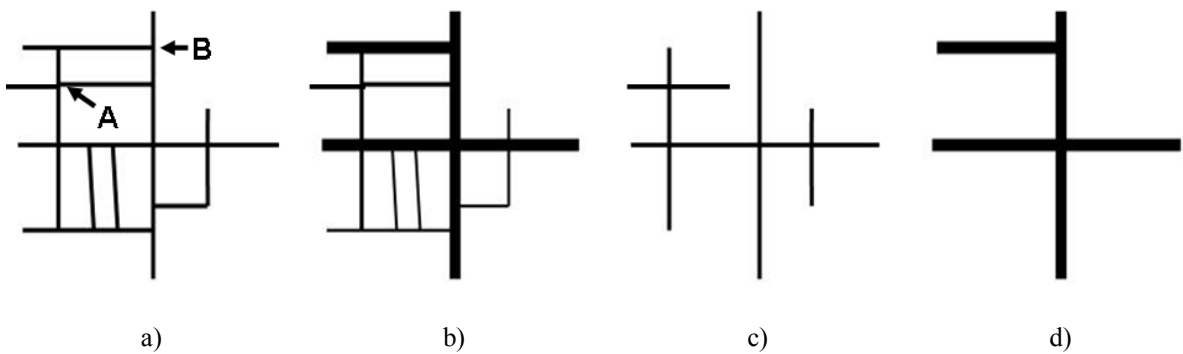

a)      b)      c)      d)

Fig. 3. Fragments of complex network (a), network system (b), and their $k$-core (c) and $\lambda$-core (d)

As already mentioned above, the real networks can consist of millions and billions elements. This significantly complicates the investigation of NS that have such structures. Solution of systems modeling problems can be simplified by studying not the entire network but its $\lambda$-core. In this case, the larger is the specific weight of $\lambda$-core in system operation, the more adequate is the result of research. It is possible to use different methods for determining the specific weight of $\lambda$-core, for example, levels of coverage of $\lambda$-core of nodes and/or edges of source network.

Let us introduce the flow adjacency matrix of $\lambda$-core $B^*_{f,\lambda} = \{b^*_{ij,\lambda}\}^N_{i,j=1}$ by the ratio

$$b^*_{ij,\lambda} = \begin{cases} b^*_{ij}, & \text{if } b^*_{ij} \geq \lambda, \\ 0, & \text{if } b^*_{ij} < \lambda. \end{cases}, \quad i,j = \overline{1,N}.$$

To determine the specific weight of $\lambda$-core, we will use parameter $\sigma^\lambda_f$, which is equal to the ratio of volume of flows that pass $\lambda$-core to the volume of flows that pass the network as a whole, i.e.

$$\sigma^\lambda_f = \sum_{i=1}^N \sum_{j=1}^N b^*_{ij,\lambda} \Big/ \sum_{i=1}^N \sum_{j=1}^N b^*_{ij}.$$

Since the main goal of existence of any NS is to provide the motion of certain type of flows [46], the parameter $\sigma^\lambda_f$ determines quantitatively the extent to which the $\lambda$-core ensures the implementation of this goal. Let us denote $\lambda_{min} = \min_{i,j=1,N} b^*_{ij}$. If $\sigma^{\lambda_{min}}_f$ is close to 1, this means that the flows are evenly distributed over network, i.e. its $\lambda_{min}$-core as a whole coincides with a source network. The closer is value $\sigma^\lambda_f$ to 1, when the value $\lambda$ is close to 1, the less part of NS where the main volume of flows is concentrated. In this case the study can be focused primarily on this part of system. In Fig. 4 uses following conventions: $V$ – volume of flows; $V_{net}$ – total volume of flows in NS; $N_{B_\lambda}$ – number of connections in network $\lambda$-core (number of non-zero elements of matrix $B^*_{f,\lambda}$); $N_B$ – total number of network connections (number of non-zero elements of matrix $B^*_f$). Network connections are ordered on the basis of increasing the volume of flows. Line 1 corresponds to the case when flows are evenly distributed along the edges of CN, lines 2-4 correspond to the cases when the flow distribution is uneven. In particular, line 4 specifies that the bulk of flows is moving in a small subnetwork of the source network.

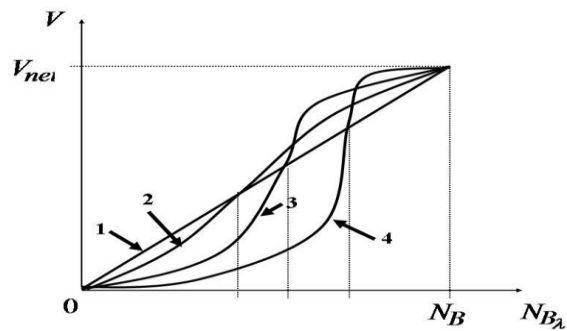

Fig. 4. Graphs of distribution of the volume of flows in $\lambda$-cores of network systems

Thus, determination and investigation of flow $\lambda$-core in compared to the study of entire NS allow us to reduce the dimension of problem by discarding nodes and edges the less important for system operation. "Disabling" these elements of NS does not lead to serious failures in its work. Consequently, research of network is reduced to analysis and mathematical modeling of most priority of its components. Cardiovascular system of human body includes both main and peripheral veins and arteries. Rupture of one of the main blood vessels can lead to rapid lethal outcome. In large city an overlap of main highways is threatening the collapse of its road transport system as a whole. Accidents on the main power lines during natural disasters often led to blackouts in entire regions of country. However, this does not mean that connections with the small values of $\lambda$ can be completely ignored. Little settlements also need to supply products, provide transportation, medical, educational and other services, in spite of the small amounts of flows. Damages of peripheral vascular leads to development of necrotic effects in organs of human body. Iterative research of $\lambda$-cores with consequent decrease of value $\lambda$ allows to simplify this process by focusing first of all on most important system components.

The problem of choice a subsystem that performs the basic functions of NS is particularly important during the modelling of large systems operation. Simplification of the model of structure can be carried out not only by means of exclusion of fictive and the least important from functional point of view nodes and edges but also the so-called nodes - mediators. The simplest example of node - mediator $n_i$ is the node with structural degree 2, which is connected to nodes $n_{i-1}$ and $n_{i+1}$ such that the elements of flow adjacency matrix $b_{i-1,i} = b_{i,i+1}$ and $b_{i+1,i} = b_{i,i-1}$. Thus, if we have some source network (Fig. 5a), then taken the 0.5-flow core of its structure (5b) and structural 3-core of this 0.5-core (5c), we obtain a substructure of NS that includes only elements important for the study of system without nodes – mediators. Such structures we will call $k(\lambda)$-cores (in Fig. 5c is reflected 3(0.5)-core of source network). It should be borne in mind that processing of flow in the node – mediator can be quite complex process [47, 48]. Destabilization of work of such node may stop the motion of all flows on the line. However, during the modelling of NS exclusion of such nodes significantly reduces the dimension of problem.

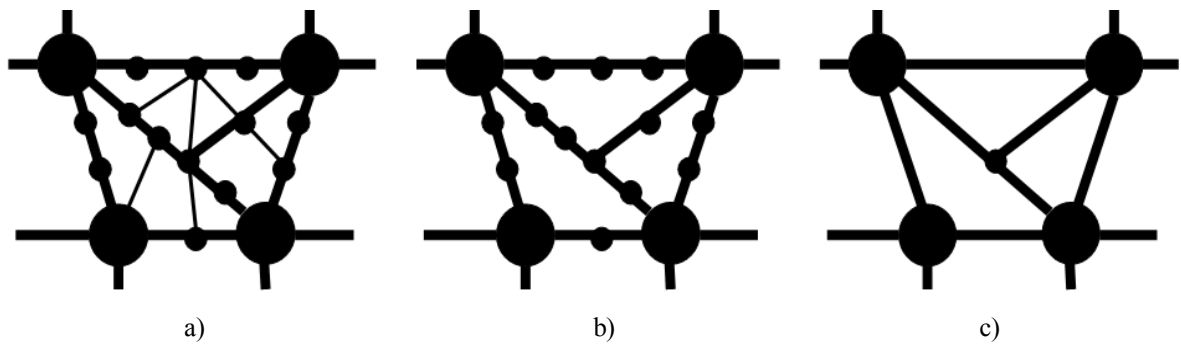

a)          b)          c)

Fig. 5. Network models with and without cores and nodes - mediators

In Fig. 6a is schematically reflected the railway network of western region of Ukraine with excluded nodes - mediators. This network contains 35 nodes and 58 edges. In Fig. 6b is reflected 0.7-core of this system, which provides more than 80% of all traffic. In Fig. 6c is reflected addition to this core, which provides less than 20% of traffic. It should be noted that 0.7-core of such system is connected network, while its addition is disconnected network. Here the dimension of 0.7-core (25 edges) is less than the dimension of its addition (33 edges). A similar phenomenon is observed in many natural and artificial network systems. Loss of connectivity of $\lambda$-core at increase of values $\lambda$ allows us to determine the network communities, the connection strength between elements of which is stronger than between other components of CN. It is obvious that the removal of part of $\lambda$-core of NS is much more destabilizing for system operation than the removal of part of its addition. Herewith, the larger part of $\lambda$-core is not available, the more problematic to find alternative paths of flows movement.

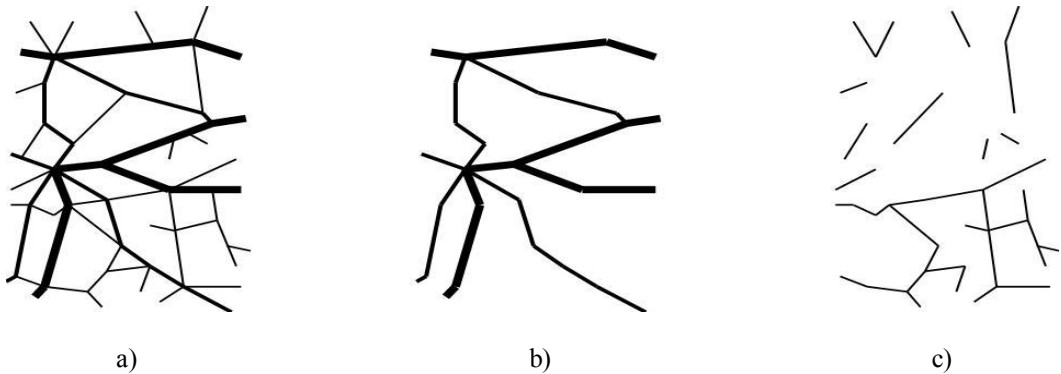

a)            b)            c)

Fig. 6. Fragment of railway network and its $\lambda$-core

In TCN the network controllability is provided by its nodes with the greatest structural degree or its *k*-core with high values *k* [35]. Controllability of NS is defined by its $\lambda$-core. It was noted above that establishment of synchronization mode depends substantially on the connection strength between network nodes. This implies that the easiest to synchronize $\lambda$-core of NS with values $\lambda$ which are close to 1. Herewith, the greater the specific weight of this core, the more its synchronization contributes to synchronization of NS as a whole. In general, the solution of problems of systems synchronization is advisable to start from its $k(\lambda)$-core. This approach is particularly convenient and justified in the case of large dimension networks. Similar reflections can be used in solving the problems of systems controllability and observability too.

Another way of defining the flow core of NS is to include in its structure only those nodes which have flow degree not less than a predetermined value. Disadvantage of this approach as the method for determining the flow core of NS is the uncertainty of determination in separate cases the connections between nodes of such core.

VII. PROBLEMS OF GROWTH AND PREFERENTIAL ATTACHMENT IN SCALE-FREE NETWORKS

In theory of complex networks the main focus is paid to study of so-called scale-free networks, which have the most implementations in reality [49]. The main feature of this type of networks, along with power distribution of structural degrees of nodes is the presence of small amounts of nodes that have a high structural degree and huge amount of nodes with a low degree (the number of metropolises in each country is small compared to the total number of settlements, but their importance in the life of country is difficult to overestimate). Two important concepts are associated with scale-free networks that describe the process of their functioning and development. The first is the concept of growth, i.e. addition of new nodes that connect with existing network nodes. The second important concept is so-called preferential attachment. It determines that more probably the new node connects with existing network nodes with high structural degree. In general, the proposed by Barabási-Albert model of development of scale-free network quite adequately describes the process of development of new territories, implementation of new technologies, expansion of infrastructure, trade, information networks, etc. At the same time, each real system after its creation is gradually moving from the stage of development to the stages of stable operation (so-called "maturity"), aging and disappearing. Flow characteristics of NS and the structure of its flow core allow us not only to develop enough adequate simplified network models, but also to solve the problems of system operation at different stages of its life cycle.

In the life cycle of any real network system we distinguish for simplification three main stages: growth or "maturation", which is accompanied by the development of structure and establishment of motion of flows; stable operation or "maturity" which is characterized by agreed motion of flows and practically unchanged network structure; curtailing or "aging" at which the volume of flows in the system is decreased and network nodes and connections between them are gradually disabled.

Growth of network (its expansion) is primarily prompted by the need to direct flows in new nodes. Conversely, the network reduction is a consequence of termination of motion of flows between existing nodes and their subsequent removal from the system structure. Barabási-Albert model is somewhat contrary to the fact that preferential attachment of new nodes in real networks more likely are performed with those nodes through which more flows are moved (Fig. 7a). Indeed, the node A with more probability connects with node B through which the motion of flows is substantially more than with node C. At the same time structural degree of node B is equal to 2, and the structural degree of node C is equal to 4. In many countries there are regions that after periods of intensive development went into state of depression (exhaustion of mineral deposits, which were mined in the region; decrease in demand for products manufactured in the region; consequences of military conflicts, etc.). In such regions usually remain developed infrastructure, in particular a dense transportation network. At the same time, the volume of flows which move such network is significantly reduced [42]. This makes unlikely connection to the nodes of such network components, regardless of their structural degree.

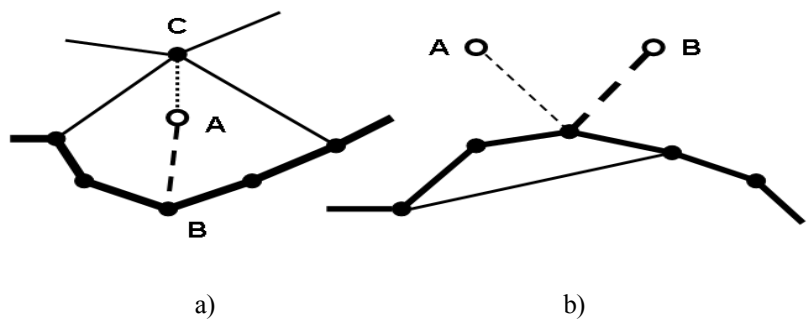

a)  b)

Fig. 7. Growth and preferential attachment in scale-free networks

As already mentioned above, the growth of scale-free networks means the connection of new nodes to already consisting in network structure. However, there is an inverse problem, which consists in determining which of two new nodes will be included in the structure at first. Accounting of motion of flows in CN gives practically definite answer to this question: the first will be included the node in which is planned to send large volumes of flows (Fig. 7b).

At the stage of "maturity" the network structure remains practically unchanged and main attention focuses on optimizing of motion of flows and improving the efficiency of the system operation as a whole.

Stage of "aging" is characterized by processes opposite to those that occurred at the stage of network growth: at first disappear the need for providing of motion of flows in certain nodes of system, and then the edges and nodes through which these flows were passed are disconnected. In many systems the first are disconnected not peripheral nodes or the nodes that were last connected but the nodes which began development of CN. An example is the development of railway system, which replaced the horse-drawn carts.

In TCN there is the concept of network density, i.e. the relationship of existing connections $L$ to all possible connections, which is calculated according to the formula $\Delta = 2L/N(N-1)$. However, not all connections that can be established between network nodes are really necessary. Moreover, to establish and maintain connections with the majority of nodes of large CN is physically impossible. For arbitrary values $N_1$, $N_2$, and $L_1$ is easy to determine the value $L_2$ such that $\Delta_1 = \Delta_2$. That is, the parameter of network density not always can track the real changes in network structure. From a systemic point of view more important is the level of coverage of the network elements by necessary flows and network capabilities to carry out such coverage.

Considering the network system as structure, which should provide the motion of certain type of flows, we can introduce a few concepts that describe the prospects for its expansion and reduction. Let us denote $N$ the number of nodes between which is organized the movement of given type of flows, and $\widetilde{N}$ is the number of nodes between which is necessary to provide the movement of these flows. Then the parameter $P_N = \widetilde{N}/N$ determines the level of coverage of network nodes of the movement of given type of flows. Let us denote $M$ the number of edges that provide the movement of given type of flows, and $\widetilde{M}$ is the number of edges that are necessary for the movement of these flows. Then the parameter $C_M = \widetilde{M}/M$ determines the level of coverage of the network nodes by connections, which are necessary for organization of movement of given type of flows. Obviously, if $P_N > 1$ and $C_M > 1$ then NS is in the growth stage. If $P_N \approx 1$ and $C_M \approx 1$, then it is in "maturity" stage or the stage of stable operation. If $P_N < 1$ and $C_M < 1$, then the system has moved to the stage of their aging and disappearing.

VIII. KERNELS AND FLOW CORES OF MULTIPLEXES

Any real system is open, i.e. it interacts with other systems [11]. One of the varieties of such interaction is multiplexes [50, 51]. Nodes of one network may be the nodes of many other networks at the same time. Thus, each country is the object (or node) of international political, economic, military, security, cultural and sports cooperation, etc. Every town in the country can be a node of several transportation networks, as well as state and local administration networks, economic and financial network. Every person is also the node of many networks (family, professional, social, religious etc.). Each network being the component of multiplex is called a layer. Examples above show that there are different types of interactions between the nodes existing on different layers of multiplex. These interactions may be of various nature or meaning and may have different material carrier. Since each network is a structure of particular network system, the same node of multiplex is an element of many systems implementing different functions therein.

During the study of multiplexes also is used the concept of **k**-core [52], as a set of *k*-cores of separate network layers. We introduce the concept of multiplex kernel, as a structure, and the flow core of multiplex as a system of interacting layers - network systems. These concepts allow us to simplify the study of systemic interactions of multiplex type.

Multiplexes are one of widespread types of interactions between the systems. Especially a lot of examples can be found in the human society. The transport system of country is familiar and understandable example of multiplex. Transport systems of various types (rail, road, air, sea and river) create a separate network layers of this structure. Banking, credit, investment and other similar structures form a financial multiplex. Trading multiplexes are combination of different commercial networks. Information multiplexes are generated by different news agencies. Linguistic (language) multiplex constitutes multilingual groups of the Earth's population. Multiplexes of social networks are combination of global (Facebook, Twitter, LinkedIn, pinterest, Google Plus, etc.) and local (Odnoklasniki, Moj Mir, VKontacte) social networks. Large technological systems are also an example of multiplex in which various physical fields interact [53].

Multiplexes are dynamic structures. Emergence of new kind of interaction between network nodes, which are included in multiplex, creates a new layer of multiplex. Conversely, if some kind of interaction disappears than corresponding layer disables too. Often a new layer and the appropriate type of interaction expands opportunities of existing multiplex layers (fixed and mobile telephony, postal service and the Internet, etc.).

From the above examples follow that in most cases the network layers of multiplex have different compositions. Therefore the multiplex is a combination of several networks with non-empty intersection the sets of their nodes (see Fig. 8).

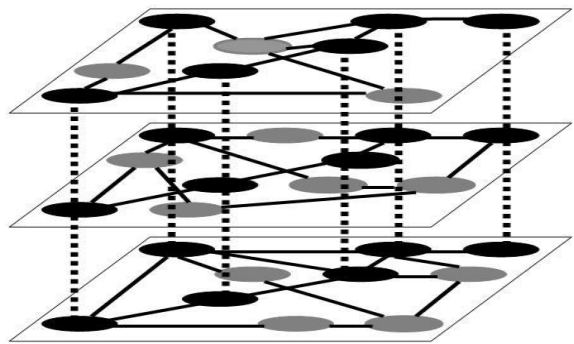

Fig. 8. Fragment of multiplex

Each network layer of multiplex reflects the structure of some network system. Consistent operation of network layer nodes contributes to achieving the goals of creation and existence of this system. Thus, multiplex is structure of supersystem of special type. This supersystem is a composition of layers - systems in each of which there is a specific character of interactions. The element (node) of each layer - system may be simultaneously an element of other layers - systems, which form the multiplex. In each of these systems it performs different functions. Such supersystems we will call the multiplex-type systems (MTS). MTS may differ by the type and nature as intralayer and interlayer interactions. In each layer is carried out one and the same character of interactions, in different layers these interactions may have a different nature. At the same time one of the main features of MTS is the availability of interlayer interactions.

MTS can be classified on the basis of possibility of flow transition without hindrance from one layer to other (passengers or cargo in transport multiplexes, information in multiplexes of social networks, news in information multiplexes, finance in banking multiplexes), compatibility of material carriers of flows, possibility of summing the volume of flows on the basis of generalized indicators (number of passengers, cargo tonnage, the total cost of all carriage, the cost of industrial or agricultural products), dependence of interlayer interactions (financial flows and industrial development, migration flows and the spread of diseases), etc.

Multiplex nodes can have different meaning both in the structure, and in the system. User can register in a dozen social networks, but do not show in them any activity or show it only in one or two. The real value of node in multiplex can be determined by its structural multiplex degree which is equal to the number of layers in which this node is included. The value of node in MTS is determined by the total volume of flows that pass through this node in all network layers.

During systems research arises many important problems: the existence of path (possibility of organizing of flow movement) from arbitrary node of one layer to arbitrary node of other layer, determining the shortest paths of movement through the multiplex (the spread of epidemics, which through multiplex interactions can escalate into pandemic, the spread of computer viruses), effect of "small world" in the multiplex, etc.

Assume that multiplex is composed of $M$ network layers. Network that consists of nodes and connections that are included in each network layer of multiplex we call a kernel of multiplex (see Fig. 9). If composition of nodes of all layers is identical and only the set of edges in each layer differ, then we get the classical definition of multiplex [51]. But, as was shown above, structure of majority of real multiplexes goes beyond such definition. The kernel of multiplex is also a complex network for study of properties of which it is expedient to use known methods of TCN.

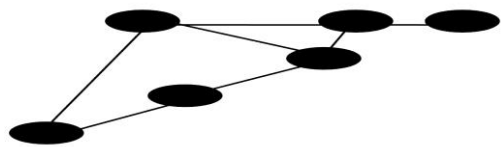

Fig. 9. Multiplex kernel

Each node of multiplex kernel is a point of transition from arbitrary network layer to any other layer. This transition can be done in different ways: the replacement of vehicle in transport multiplexes, distribution of information from one social network among other networks, communication in different language environments (network layers of linguistic multiplex), etc.

Let us denote $B_{f,K_M}^{*,m} = \{b_{ij}^{*,m}\}_{i,j=1}^{N_m}$ and $B_{f,K_M}^{*,\ker,m} = \{b_{ij}^{*,\ker,m}\}_{i,j=1}^{N_m}$ the flow adjacency matrices for *m*-th layer and multiplex kernel in *m*-th layer, respectively, $N_m$ is a number of nodes in *m*-th layer, $m = \overline{1,M}$. Specific weight $q_{K_M}^{(m)}$ of multiplex kernel in *m*-th network layer is determined by the ratio

$$q_{K_M}^{(m)} = \sum_{i=1}^{N_m}\sum_{j=1}^{N_m} b_{ij}^{*,\ker,m} / \sum_{i=1}^{N_m}\sum_{j=1}^{N_m} b_{ij}^{*,m}.$$

The larger specific weight of multiplex kernel in *m*-th layer, the more its properties determine the properties of this layer. Furthermore, the relative number of transition points in the layer is important for organizing interlayer system interactions.

Various network layers may have a different priority in the system. This priority can be determined by branching of nodes and edges of the layer (for example, the branching of road network is greater than rail network, rail network - more than aviation network). Priority of layers can be determined by a comparative analysis of total volume of flows that are moving in these layers (financial flows in one of banking networks can be grater than in other). In the case of different priority of layers the specific weight of kernel in MTS $q_{K_M}^{wla}$ we will calculate using the method of weighted linear aggregation (MWLA), which is determined by the ratio [54]

$$q_{K_M}^{wla} = \sum_{m=1}^{M} \rho_m q_{K_M}^{(m)} / \sum_{m=1}^{M} \rho_m,$$

where $\rho_m$ is weight coefficient which determines the priority of *m*-layer, $m = \overline{1,M}$.

If network layers have the same priority, the specific weight of kernel $q_{K_M}^{na}$ in MTS we will calculate using nonlinear aggregation method (NAM), which is determined by the ratio [55]

$$q_{K_M}^{na} = \prod_{m=1}^{M} q_{K_M}^{(m)} / (q^*)^{M-1},$$

where $q^* = \sum_{m=1}^{M} q_{K_M}^{(m)})/M$. Note that in the case of the same priority of layers NAM gives a more accurate estimate of specific weight of kernel in MTS than WLAM.

If the network layers of multiplex can be divided into groups with different priority (layers of group have the same priority), the specific weight of kernel in MTS is determined by method of hybrid aggregation [56] that combines the advantages of WLAM and NAM. Obviously, the greater specific weight of kernel in MTS, the greater its properties define properties of the system as a whole.

In [52] **k**-core of multiplex $\mathbf{k} = \{k_1, k_2, ..., k_M\}$ is defined as the largest subgraph in which each node has at least $k_m$ edges of separate type, $m = \overline{1,M}$. In each network layer of multiplex we can define its flow core. Flow core of MTS can be determined if the flows in all network layers are of the same type (passengers or cargoes), or if the flows are polytypic, but there is a general indicator of their movement, for example, the cost of transportation (in every country there is statistics of annual passenger and cargo traffic, as well as summary measures of the cost of transportations). Then the flow core of MTS is defined as the flow core of multiplex kernel with total indicators of flow interactions (volume of flows movement), which can not be less than a predetermined value. It is obvious, that the kernel of multiplex is a structural concept and the flow core of MTS is a system concept.

Determination of specific weight of flow core of MTS in each layer, the kernel of multiplex and MTS as a whole is similar to the definition of specific weight of multiplex kernel in each layer and a whole multiplex. In this case, the greater the specific weight of flow core in each layer, kernel and MTS, the greater its properties determine the properties of corresponding structure or system. It is obvious that the preferential attachment of layer nodes are more likely carried out with the kernel nodes, as it gives possibility of fast transition to other layers of multiplex. Among the kernel nodes the preferential attachment is more probable to nodes of flow core of MTS.

In MTS can much more effectively solve the problems of search the alternative paths of movement. The possibility of movement of flow through kernel nodes from one layer to another and vise versa allows us to move it between disconnected components of separate network layer (see Fig. 2c). Examples of such transition are the use of subway in large cities or the replacement of rail transport by road or air. On the other hand, the unavailability of part of the kernel or core of MTS is substantial threat for its functioning. A separate issue is the phenomenon of "small world" in MTS. It is also interesting are the problems of synchronization, controllability and observability of interlayer interactions, etc.

## IX. CONCLUSIONS

The theories of complex networks and network systems are in developing state now. They still require the development of conceptual apparatus, formulation of problems that need to be solved, extension the applied aspects of use. In this article we considered only some of the basic concepts and problems of complex network theory and introduced their flow analogues. Obtained results evidence that the accounting of flows movement in the network can not only enrich the problematic of TCN, but also to bring it closer to the problematic of systems theory and systems analysis: to classify network systems, identify the goals of their creation and development, determine functions that contribute to the achievement of these goals, to solve controllability problems, etc. Definition of local and global flow characteristics of the network allows for each concrete case to build a flow model of real network system and to solve a set of important problems that are now difficult to solve using only features of TCN: identification of fictive and hidden nodes and edges, determining the actual level of network clustering, structure of cliques, communities and other its substructures, importance of hubs, etc.

Concept of flow core of network system allows us together with the concept of network $k$-core substantially reduce the dimension of problems of modeling the behavior of CN and NS. Flow characteristics of network systems make it possible to more realistically determine the priority directions of development and preferential attachment in scale-free networks and investigate the processes of functioning of NS at all stages of their life cycle. Concept of multiplex kernel allows us to determine the transition points between all network layers and optimize the search of shortest paths of flows movement. Methods for determining the specific weight of kernel in each layer and the whole multiplex are proposed. Concept of flow core of multiplex type systems is defined and possibility of its construction depending on the types of flows in different network layers is considered. Methods for determining the specific weight of flow core in separate layer, kernel and the whole multiplex are proposed. Introduced concepts significantly simplify the investigation of complex network systems and multiplex type systems. At the same time, the larger is the specific weight of core and kernel in corresponding structures, the more adequate is the result of research.